\documentclass[prb,aps]{revtex4}
\usepackage{epsfig}
\begin{document}

\title{Broken Time Reversal Symmetry and Superconducting States in the Cuprates}
\author{ R.P. Kaur and D.F. Agterberg}
\address{Department of Physics, University of Wisconsin-Milwaukee, Milwaukee, WI 53211}

\begin{abstract}
Recently, Kaminski {\it et al.} have reported that time reversal
symmetry is broken in the pseudogap phase in the high temperature
superconducting material Bi$_2$Sr$_2$CaCu$_2$O$_{2+\delta}$
(Bi-2212). Here we examine the role of translationally invariant
broken time reversal states on $d_{x^2-y^2}$  superconductors. In
particular, we determine the change in the superconducting order
parameter structure. We find that the broken time reversal
pseudogap state that is consistent with the experiment of Kaminski
{\it et al.}, gives rise to a novel mixed singlet-triplet pairing
$d+ip$ phase. This $d+ip$ state is shown to give rise to a helical
superconducting phase. Consequences of this $d+ip$ state on
Josephson experiments are discussed.
\end{abstract}

\maketitle

\begin{figure}
\epsfxsize=5.0 in \center{\epsfbox{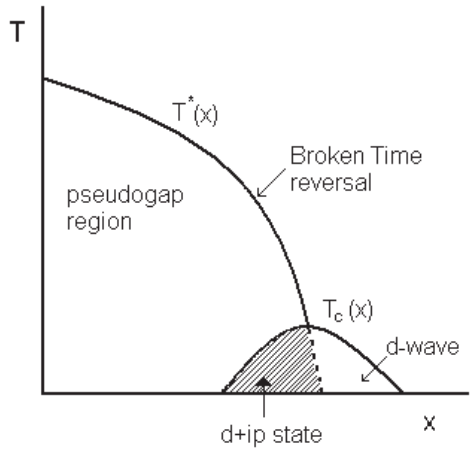}}
\caption{Possible phase diagram, temperature T as a function of
hole doping, of the cuprate Bi-2212. Here $T^*$ is the critical
temperature where time reversal symmetry is broken and $T_c$ is
the transition temperature of superconducting phase.} \label{fig1}
\end{figure}

 The origin of the pseudogap regime in the cuprates has been the subject of controversy. A variety of probes reveal
 a suppression of the single particle density of states in this regime \cite{din96,loe96}.  A natural
 explanation is that the pseudogap phase is a precursor superconducting state; a phase in which there
 are Cooper pairs but no superconducting phase coherence \cite{ran92}. The recent experimental results of Kaminski
 {\it et al.} \cite{camp} and of Alff {\it et al.} \cite{welter} provide evidence for a very different
 explanation of the pseudogap phase:
 it marks a new phase in which a symmetry is broken. In particular, the results of Ref.~\onlinecite{camp}
 have reported that left circularly polarized photons give a different
 photocurrent from right-circularly polarized photons in the
 pseudogap phase. This, combined with a mirror plane symmetry, implies
 the breaking of time reversal symmetry in the pseudogap phase.
 Varma \cite{varma1} has
 proposed translationally invariant orbital current states that may account for the observed results.
 If it is indeed the case that the pseudogap phase does not break translational symmetry and breaks time reversal
 symmetry, then the classification of the superconducting pairing symmetry will differ from previous classifications.
 Fig.~\ref{fig1} shows the resulting phase
 diagram. In this paper, we determine the structure of the superconducting
gap function in the pseudogap phase when time
 reversal symmetry is broken and translational symmetry is preserved. Note that this implies that we do not consider the d-density wave
 state of Chakravarty {\it et al.} \cite{supid1}. The reason for this restriction is that a pseudogap phase which breaks
 translationally symmetry will not alter the superconducting state as strongly as the case considered here.
 The layout of the paper is as follows:
 we first find the possible superconducting states in the pseudogap
 phase by using both corepresentation theory and phenomenological Ginzburg Landau arguments. This is done for all
 pseudogap symmetries that retain translational symmetry.
Then we focus on the pseudogap phase that is consistent with the
experimental results of Kaminski {\it et al.} This phase is
determined by the requirement that the pseudogap phase breaks time
reversal symmetry, breaks
 the four-fold rotation symmetry of the CuO$_2$ plane, and also breaks
 the mirror plane symmetry with normal along the Cu-O diagonal. This leads to the two possible pseudogap order parameters
 that have been examined by Varma and Simon \cite{varma1,varma2}. A detailed symmetry analysis of the
 photoemission matrix elements rules out one of these two order parameters \cite{varma2}.
We develop a Ginzburg-Landau theory of the remaining pseudogap
order parameter and the interplay of of this order parameter with
superconductivity. We
 discuss some observable consequences of the superconducting gap
 components induced by the pseudogap order parameter. In particular, we
show that a helical superconducting phase is a consequence of
broken time reversal symmetry in this case. Finally we discuss the
$c$-axis Josephson current that has been observed in strongly
underdoped Bi-2212 - Pb junctions \cite{Jose}. We show how this
can possibly be explained by the coupling of the superconducting
order parameter to domain walls in the pseudogap order.



Here we describe the possible magnetic point groups of the
pseudogap phase and their respective free energies. The symmetry
group of the pseudogap phase can be written as $\cal G$=$
G_M\times U(1)$ where $G_M$ is the magnetic space group and $U(1)$
is the gauge group (which is not broken in the pseudogap phase).
The group $G_M$ is the group that leaves both the charge density
and the magnetization density ${\bf M}$ invariant. We will be
interested in the possible magnetic point groups that arise when
time reversal symmetry is broken in Bi-2212. We will focus only on
transitions that do not break translational invariance and thus
focus on the $4/mmm$ ($D_{4h}$) point group which is defined
through the elements $\{ E,C_{2x}, C_{2y}, C_{2z}, i, \sigma_{x},
\sigma_{y}, \sigma_{z}, C_{da}, C_{db}, \sigma_{da}, \sigma_{db},
\pm C_{4z}, \pm i C_{4z} \}$. Magnetic point groups are defined as
$S_M= H+\theta(G-H)$, where $H$ is a halving subgroup of the
ordinary point group $G$. In magnetic symmetry groups, the crystal
point group operation $R$, which belongs to $G-H$ group,
transforms the magnetization density ${\bf M}$ to $-{\bf M}$, but
the product of the time reversal operation $\theta$ and the
operation $R$ leaves ${\bf M}$ invariant. In Table \ref{Table1}, a
list of magnetic point groups corresponding to pseudogap phase has
been given. According to the experimental observations of Kapinski
{\it et al.} \cite{camp}, the four fold rotation about z-axis
($C_{4z}$) and the diagonal mirror planes ($\sigma_{da},
\sigma_{db}$), are no longer symmetry operations in the pseudogap
state. The requirement that these two symmetries are no longer
present reduces the number of possible magnetic point groups of
the pseudogap phase to two: $\underline{4}/mm\underline{m}$ $\{ E,
C_{2x}, C_{2y}, C_{2z}, i, \sigma_{x}, \sigma_{y}, \sigma_{z},
\theta C_{da}, \theta C_{db}, i \theta C_{da}, \theta \sigma_{db},
\pm \theta C_{4z}, \pm i \theta C_{4z} \}$ or $\underline{m}mm $
$\{E, C_{da}, \sigma_z, \sigma_{db}, i \theta , \theta
\sigma_{da}, \theta C_{2z}, \theta C_{db} \}$. The symmetries,
$\underline{4}/mm\underline{m}$ and $\underline{m}mm $ agree with
the proposed orbital current patterns of Ref.~\onlinecite{varma1}
\cite{varma3}. Consequently, we use the same notation: we label
the group $\underline{4}/mm\underline{m}$ as type I and the group
$\underline{m}mm$ as type II. A detailed symmetry analysis of the
photoemission matrix elements rules out the type I state. So we
will consider the type II phase in more detail (note that Stanescu
and Phillips have also examined the type II phase microscopically
\cite{phillips}).

For the one dimensional pseudogap order parameters (A or B
irreducible representations (REPS) of the $D_{4h}$ point group in
Table~\ref{Table1}), the Ginzburg Landau free energy is simply
given as
\begin{equation}
F_{pg,I}[\eta]=\alpha_{1} \eta^2 + {\beta_{1}\over 2}\eta^4.
\end{equation}
Two degenerate states $\eta=\pm \sqrt{-\alpha_1/ \beta_1}$
minimize this free energy.

For the two dimensional pseudogap order parameters ($E_u$ or $E_g$
REPS of $D_{4h}$ point group in Table~\ref{Table1}),
the corresponding Ginzburg Landau free energy is
\begin{equation}
F_{pg,II}[\eta_x,\eta_y]=\alpha (\eta_x^2+\eta_y^2)+{\beta \over
2} (\eta_x^4+\eta_y^4)+ \gamma \eta_x^2 \eta_y^2. \label{FreeII}
\end{equation}
Minimization of the free energy with respect to $\eta_x$ and
$\eta_y$ gives the following sets of degenerate states:
\begin{equation}
(\eta_x,\eta_y) =(0,0); \sqrt{-\alpha \over \beta}\{(0,\pm1),
(\pm1,0)\}; \sqrt{-\alpha \over \beta+\gamma}\{(\pm1,\pm1),(\pm1,
\mp1)\}.
\end{equation}
Of these states, $\eta_0(\pm1,\pm1)$ and $\eta_0(\pm1,\mp1)$,
where $\eta_0=\sqrt{-\alpha/(\beta+\gamma)}$, minimize the free
energy if $\gamma<\beta$. The other four degenerate states
$\acute{\eta}_0(0,\pm 1), \acute{\eta}_0(\pm 1,0)$, where
$\acute{\eta}_0=\sqrt{-\alpha/\beta}$, minimize the free energy
when $\gamma>\beta$. Simon and Varma have depicted the $(1,1)$
state to represent the type II current pattern \cite{varma1}, but
there are three more states which are degenerate with this state.
In these states, the tetragonal symmetry is lost as
$C_{4z}(1,1)=(1,-1)$. Each domain has symmetry $\underline{m}mm$
with the $\underline{m}$ oriented along the different diagonals.
The $\sigma_{da(db)}$ symmetry will be lost if there are multiple
domains. Note that the observed incommensurate modulation in
Bi-2212 implies the existence of an $\epsilon_{xy}$ strain
\cite{Jose}. This gives rise to an additional
$\epsilon_{xy}\eta_x\eta_y$ invariant in the pesudogap free
energy. This prefers the $(\pm1,\pm1)$ and $(\pm1,\mp1)$ states
and also breaks the degeneracy of these two states so that they
are now two-fold degenerate.


To classify the possible superconducting states in the presence of
broken time reversal symmetry requires the use of corepresentation
theory. The usual representation theory must be extended because
the time reversal operator is antilinear and antiunitary. The
magnetic point group can be written as $S_M=H+A\times H$, where
$A$ is antiunitary operator such that all elements of the coset
$AH$ are antiunitary. The corepresentations (COREPS) $D\Gamma$ of
$S_M$ can be found from REPS $\Gamma$ of the corresponding normal
group $H$ in one of three ways [labelled (a), (b), or (c)]. This
approach is explained in Ref.~\onlinecite{book} and we use their
notation. Recently, similar considerations have appeared in the
classification of superconducting states in ferromagnets
\cite{fomin,walker,mineev}. The superconducting gap function is
defined as \cite{sig91,gor87} $\hat{\Delta}({\bf k})=i(\psi({\bf
k})+i {\bf d}({\bf k}).\hat{\sigma})\hat{\bf \sigma}_2$. In Table
~\ref{Table1} we give representative gap functions $\psi({\bf k})$
and ${\bf d}({\bf k})$ for various pseudogap symmetries.

As an example, consider the $\underline{4}/mm\underline{m}$
magnetic point group. The pseudogap order parameter in this case
corresponds to a one dimensional REP of $D_{4h}$. For this
magnetic point group, $H=D_{2h}$ and $A=\theta {C_{4z}}$. For a
$d_{x^2-y^2}$ pairing symmetry when time reversal symmetry is not
broken, we are interested in the $A_{1g}$ REP of $D_{2h}$. The
resulting pairing state is a real linear combination of $\psi({\bf
k})=k_x^2+k_y^2$ and $\psi({\bf k})=i(k_x^2-k_y^2)$ which can be
denoted as a $d+is$ pairing state


\begin{table}[h]
\begin{center}
\begin{tabular}{|c|c|c|}\hline
Pseudogap   & Induced & Basis functions\\
Phase\scriptsize{(REP)} &\scriptsize{(REP)} & \scriptsize{(COREP)}\\
\hline \hline
 $A_{1g}(4/mmm)$       & $B_{1g}$ & $(k_x^2-k_y^2)[1+ia(k_x^2+k_y^2)]$\\\hline
 $A_{2g}(4/m\underline{mm})$   & $B_{2g}$ & $(k_x^2-k_y^2)+iak_x k_y $\\\hline
 $B_{1g}(\underline{4}/mm\underline{m})$   & $A_{1g}$ & $(k_x^2-k_y^2) + ia(k_x^2+k_y^2)$\\\hline
 $B_{2g}(\underline{4}/mm\underline{m})$   & $A_{2g}$ & $(k_x^2-k_y^2)+iak_x k_y(k_x^2-k_y^2)$\\\hline
 $E_{g}(m\underline{mm})$ & $E_{g}$  & $(k_x^2-k_y^2)+ia(\eta_1k_y-\eta_2 k_x)k_z$\\\hline
 $A_{1u}(4/\underline{mmm})$ & $B_{1u}$ & $(k_x^2-k_y^2) +ia(\hat{\sigma_1}k_x-\hat{\sigma_2}k_y)$\\\hline
 $A_{2u}(4/\underline{m}mm)$     & $B_{2u}$ & $(k_x^2-k_y^2)+ia(\hat{\sigma_1}k_y+\hat{\sigma_2}k_x)$\\\hline
 $B_{1u}(\underline{4}/\underline{m}m\underline{m})$ & $A_{1u}$ & $(k_x^2-k_y^2)+ia(\hat{\sigma_1}k_x+\hat{\sigma_2}k_y)$\\\hline
 $B_{2u}(\underline{4}/\underline{m}m\underline{m})$ & $A_{2u}$ & $(k_x^2-k_y^2)+ia(\hat{\sigma_1}k_y-\hat{\sigma_2}k_x)$\\\hline
 $E_{u}(\underline{m}mm)$ & $E_{u}$  & $(k_x^2-k_y^2)+ia(\eta_1k_y-\eta_2k_x)\hat{\sigma_3}$\\\hline
\end{tabular}
\end{center}
\caption{Superconducting COREP basis functions for different
magnetic point group symmetries of the pseudogap phase. Induced
REP refers to the superconducting component induced by the
pseudogap order. The magnetic point groups for two dimensional
pseudogap REPS depend upon the form of $\eta_1$, $\eta_2$, here we
take $\eta_1=\pm\eta_2$.} \label{Table1}
\end{table}

The magnetic point group $\underline{m}mm$ deserves further
consideration since it is consistent with the experimental
observations of Kaminski {\it et al} \cite{varma2,camp}.
For this group, $H=C_{2v}$ and $A=i\theta$. Due to the broken
parity symmetry in the pseudogap phase, the pairing gap function
is a mixture of spin-singlet [$\psi({\bf k})$] and spin-triplet
components [${\bf d}({\bf k})$].  For a $d_{x^2-y^2}$ pairing
symmetry when time reversal symmetry is not broken, we are
interested in the $B_{1}$ REP of $C_{2v}$ (given in
Table~\ref{Table2}). The corresponding COREP is a real linear
combination of the spin-singlet $\psi({\bf k})=(k_y^2-k_x^2)$ and
the spin-triplet ${\bf d}({\bf k})=i(k_x-k_y) \hat{z}$ gap
functions. We label this state the $d+ip$ phase. In Table II, we
have also provided representative basis functions for the other
COREPS corresponding to the different REPS of $C_{2v}$. From the
form of the gap functions found here we can deduce whether there
are possibly nodes in the gap.  We find that the superconducting
gap will vanish along the $k_x=k_y$ line for in this case (the gap
does not vanish along the $k_x=-k_y$ line). Note that similar
considerations appear in a recent symmetry analysis of a d-wave
superconductor in a uniform current \cite{kabanov}.


\begin{center}
\begin{table}
\begin{tabular}{|c|c|c|c|c|c|}\cline{1-6}
& \shortstack{E\\ $i \theta$} & \shortstack{\rule{0mm}{1mm}\\ $C_{da}$ \\
$\theta\sigma_{da}$} & \shortstack{$\sigma_z$ \\ $\theta C_{2z}$}
& \shortstack{$\sigma_{db}$
\\ $\theta C_{db}$} & \shortstack{$\psi({\bf k})+
 {\bf d}({\bf k}).\hat{\sigma}$\\ Basis{\scriptsize(COREP)}} \\ \hline \hline
 {$A_1$}& 1 & 1 & 1 & 1 &  $(k_x^2+k_y^2)+a k_x k_y+ib\sigma_3k_{da}$ \\ \hline
 {$A_2$}& 1 & 1 & -1 & -1 & $(k_{db}k_z+ia(\sigma_1 k_x+\sigma_2 k_y)$ \\
      &   &   &    &     & $+ib(\sigma_1 k_y+\sigma_2 k_x)$\\ \hline
 {$B_1$}& 1 & -1 & 1 & -1 &  $(k_y^2-k_x^2)+ia\sigma_3k_{db}$ \\ \hline
 {$B_2$}& 1 & -1 & -1 & 1 &  $k_{da}k_z+ia(\sigma_1 k_y-\sigma_2 k_x)$ \\
        &   &    &    &   & $+ib(\sigma_1 k_x-\sigma_2 k_y)$\\ \hline
 \end{tabular}
\caption[Table 2.]{The COREPS of $\underline{m}mm$ group with
$H=C_{2v}$. In the Table, $a$ and $b$ are arbitrary constants,
$k_{da}=k_{x}+k_y$, $k_{db}=k_x-k_y$, and $\sigma_1, \sigma_2$ and
$\sigma_3$ are Pauli matrices. The COREPS are all type (a).}
\label{Table2}
\end{table}
\end{center}

Here we determine the superconducting gap structures in the
pseudogap phase using Ginzburg Landau theory. This approach
provides the same results as those found above using the less
familiar corepresentation theory. Using corepresentation theory,
it has been shown that the existence of $d$-wave order parameter
$\psi_d$ and pseudogap order parameter $\eta$ ensures the
existence of an induced superconducing order parameter $\psi$. In
Table \ref{Table1}, the different possible combinations of order
parameters of  the pseudogap order $\eta$ and of induced
superconducting order $\psi$ have been listed. These same states
can be found by examining Ginzburg Landau theory. In particular,
the invariance of the free energy with respect to time reversal
symmetry requires any free energy invariant corresponding to any
one dimensional pseudogap REP to have the form $i \epsilon \eta
(\psi_{d}^*\psi-\psi^*\psi_{d})$, where $\epsilon$ is real
coupling coefficient. Therefore, the superconducting Ginzburg
Landau free energy corresponding to the one dimensional pseudogap
REPS of Table \ref{Table1} (up to second order is in $\psi$) is :
\begin{equation}
F_I(\eta,\psi_{d},\psi)=F_{pg,I}+ \alpha_{d} |\psi_{d}|^2
+\beta_d|\psi_d|^4+ \tilde{\alpha} |\psi|^2 +
i\epsilon\eta(\psi_{d}^{*}\psi-\psi^{*}\psi_{d})\label{FgI}
\end{equation}
The relation between the order parameters can be obtained by
minimizing $F(\eta,\psi_{d},\psi)$ with respect to $\psi^*$. The
relation is $\psi=\pm i \epsilon \eta \psi_{d} /\tilde{\alpha}$,
where $\pm$ sign is due to the degeneracy of $\psi$ and $-\psi$
states. From the relation it is clear that for the non-zero $\eta$
and $\psi_{d}$ order parameters, $\psi$ must be non-zero, which
ensures that the symmetry of superconducting state in pseudogap
phase is $d+i\psi$.

For a type II pseudogap phase, the order parameter belongs to the
two dimensional $E_u$ REP of the $D_{4h}$ point group. Since the
product of the d-wave order parameter $\psi_d$ and pseudogap order
parameter $(\eta_x,\eta_y)$ transforms as a $E_u$ REP, the induced
superconducting order parameter also belongs to the $E_u$ REP. The
induced  superconducting order parameter can be written as
$\psi=(p_x, p_y)$. To construct a non-trivial invariant, we
decompose the product of representations $E_u \bigotimes E_u
\bigotimes B_{1g}$.  The relevant invariant is $i \tilde{\epsilon}
[\psi_d^* (\eta_x p_x -\eta_y p_y)-\psi_d (\eta_x p_x^* -\eta_y
p_y^*)]$, where $\tilde{\epsilon}$ is a real positive coefficient.
The Ginzburg Landau free energy [up to second order in
$(p_x,p_y)$] is
\begin{equation}
F_{II}(\psi_{d},\eta,p)=F_{pg,II}+\alpha_{d} |\psi_{d}|^2 +
\beta_d|\psi_d|^4+\alpha_{p} (|p_x|^2+ |p_y|^2) +
i\tilde{\epsilon} [\psi_d^* (\eta_x p_x -\eta_y p_y)-\psi_d
(\eta_x p_x^* -\eta_y p_y^*)].\label{FgII}
\end{equation}
Minimizing the free energy with respect to $p_{x}^{*}$ and
$p_{y}^{*}$ gives $p_x=i \tilde{\epsilon} \psi_{d}
\eta_{x}/\alpha_{p}$ and $p_y=-i \tilde{\epsilon} \psi_{d}
\eta_{y}/\alpha_{p}$. Thus we have $(p_x,p_y)=i \tilde{\epsilon}
(\eta_x, -\eta_y) \psi_d/\alpha_{p} = i\tilde{\epsilon}
\eta_0(1,-1) \psi_d/\alpha_{p}$ where
$(\eta_x,\eta_y)=\eta_0(1,1)$ has been used.\\

The appearance of the induced superconducting gap functions in the
pseudogap phase leads to a variety of observable consequences.
In the pseudogap phase having $d+ip$ state, the breaking of parity
symmetry gives rise to Lifshits invariants in the free energy
which give rise to a spatially varying (helical) superconducting
phase \cite{min94}. This behavior can be readily explained in
terms of the Ginzburg Landau free energy. In particular, the
relevant Lifshitz invariant is
\begin{equation}
F_L=F_{II}+i\acute{\epsilon} [\psi_d^* (D_x p_x -D_y p_y)-
\psi_d(D_x^* p_x^* -D_y^* p_y^*)].\label{F1d+ip}
\end{equation}
where ${\bf D}=(D_x,D_y)$, $D_j=- i \nabla_j-2eA_j/\hbar c$, and
$\bf A $ is the vector potential. The helical superconducting
phase can be found by setting ${\bf A}=0 $ and considering the
spatial variation of order parameters as
$\psi_d=\psi_{d0}e^{i\bf{q.r}}$ and
$(p_x,p_y)=(p_{x0},p_{y0})e^{i\bf{q.r}}$. Minimizing with respect
to $q_x$ and $q_y$ gives $q_x=q_y=-{\acute{\epsilon}\epsilon
\eta_0 \over \alpha_p\tilde{ \kappa}}$ (where $\tilde{\kappa}$ is
the defined through the gradient term $\tilde{\kappa}|{\bf
D}\psi_d|^2$ in the free energy). Note that gauge invariance and
minimization of the free energy with respect to ${\bf q}$ implies
that the current in helical phase is zero. The helical structure
of the order parameter can be verified by Josephson junction
experiments. We refer to Ref.~\onlinecite{yang} for details. The
existence of gap nodes found in the last section assumed a uniform
(non-helical) order parameter. If $1/q_x>>\xi_0$, where $\xi_0$ is
the coherence length, then the nodes will presumably still provide
a reasonable description of the low energy excitations of the
superconductor.

It is interesting to note that a Josephson current has been
observed through $c$-axis Josephson junctions between {\it
underdoped} Bi-2212 and Pb. Mo$\ss$le {\it et al.} have
demonstrated this through a series of very careful experiments
\cite{Jose}. They have shown that the junctions are homogeneous
and the coupling is a conventional lowest-order Josephson
coupling. This experiment is difficult to explain with a pure
$d$-wave order parameter in Bi-2212. Rae has pointed out that this
current may exist if the Pb superconductor contains a $d$-wave
contribution \cite{rae}. This is possible if the junction is made
from a low-symmetry orientation of lead (a [110] face, for
example). However, as pointed out by Rae, this explanation
requires that there is a systematic bias towards [110] Pb faces in
the junctions which remains to be verified. Given the possibility
of a $d+ip$ phase, it is natural to ask if this can be related to
the observed Josephson current. Here we show that while a $d+ip$
state as described above does not have a Josephson current; a
domain wall in the pseudogap order parameter will. This can be
understood by considering surface energy at a $c$-axis junction
between a $d+ip$ superconductor and a conventional $s$-wave
superconductor;
\begin{equation}
F_{sur}=\int d^2S[\psi_s^*(\eta_x p_x+\eta_y p_y)+c.c.]
\end{equation}
If we consider $(\eta_x,\eta_y)=\eta_0(1,1)$ then, as shown above,
$p_x=-p_y$ and $F_{sur}$ will be zero. But consider a domain wall
of the type, $(\eta_x,\eta_y)=\eta_0[1,\tanh(x/\xi)]$(here
$\eta_x$ remain constant but $\eta_y$ varies from -$\eta_0$ to
$\eta_0$ across the domain wall), the current density in this case
is given as:
\begin{equation}
j= j_0[1-\tanh^2(x/ \xi)]\sin(\theta_d-\theta_s),
\end{equation}
where $j_0=4e\eta_0^2\tilde{\epsilon}|\psi_d||\psi_s|$. A
Fraunhofer pattern can be obtained by applying the field along the
normal to the domain walls. However, as the angle between the
field and domain walls decreases, the current pattern will deviate
from the usual Fraunhofer pattern. In particular, for the field
along the domain wall, only the central peak of the original
Fraunhofer pattern remains. Note that these considerations require
the junction size to be much smaller than the period of the
helical order, so that the $p$-wave order parameter is
approximately uniform. Furthermore, in Bi-2212, the incommensurate
modulation may alter this analysis. If there is a strong
interaction between this modulation and the pseudogap order
parameter, then any domain walls in the pseudogap order parameter
will be tied to domain boundaries of incommensurate modulation.

In conclusion, we have determined the superconducting gap
structure in translationally invariant pseudogap phases that break
time reversal symmetry. It has been shown that a $d+ip$ state is
the superconducting ordered state for the pseudogap state that
agrees with experimental results in Bi-2212. The induced $ip$
component removes two of the four nodes associated with a
$d_{x^2-y^2}$ order parameter. The consequences of the induced
$ip$ order parameters on Josephson experiments has been explored.
The induced $ip$ phase can explain the observed Josephson current
through a $c$-axis junction between underdoped Bi-2212 and Pb only
if there are the domain walls in the pseudogap order parameter. It
also has been shown that the $d+ip$ state will give rise to a
helical superconducting phase.

We acknowledge the Donors of the American Chemical Society
Petroleum Research Fund for support of this work and we thank V.V.
Kabanov for useful correspondence.

\end{document}